# Validity of Non-Hermitian fluctuation theorem in open Quantum system with unbroken PT-symmetry

Saurov Hazarika

*Department of Chemistry, Johns Hopkins University, Baltimore-21218, USA*

E-mail: shazari1@jhu.edu.in

**Abstract**

In a recent paper, Deffner and Saxena (2015 *Phys. Rev. Lett. 114* 150601) showed that quantum Jarzynski equality generalizes to $\mathcal{PT}-$ symmetric quantum mechanics with unbroken PT symmetry. Later Zeng and Yong (2017 *Journal of Phys. Commun. 1* 031001) extended this work to Crook's fluctuation theorem. In another recent paper, Andrew Smith *et al* 2018 discusses non-equilibrium work relation in open quantum system. In this paper, we will discuss the validity of Non- Hermitian fluctuation theorem in open quantum system, in a region of unbroken PT-symmetry.

## 1. Introduction

In recent time, the far from equilibrium phenomena have attracted a lot of attention in the scientific community. In 1993 Evans *et al.* and in 1994 Gallavotti *et al.* have developed a series of fluctuation theorems, making the earliest breakthroughs in the field. In his seminal paper [1], Jarzynski derived a very fundamental relationship between non-equilibrium work and free energy difference. The relation reads as

$$\langle e^{-\beta W}\rangle = e^{-\beta \Delta F}$$

here W stands for work done when a driving force acts on the system and drives the system out of the equilibrium, and ΔF stands for the change in free energy during the process. In 1999, Gravin E Crooks, in his papers [3,4], extended this relation and derived a more powerful relationship which is known as Crook's fluctuation theorem. The theorem states

$$\frac{P_F(+\beta W)}{P_R(-\beta W)} = e^{\beta(W-\Delta F)}$$

In recent years, a lot of works have been done in generalizing these result for the Quantum system [5,9-11]. But, although a remarkable progress has been made in discussing fluctuation theorems for closed quantum systems, there are many questions yet to discussed for Quantum Open system. In 2013, Rastegin [17] has discussed these non-equilibrium equalities with unital quantum channels. In 2017, Smith *et al*. [19] discussed the experimental verification of it. In their paper, Smith *et al*. considered the correlation between system and bath very weak, so that while decoherence plays an important role in the process, dissipation can be neglected.

In 1998, C. M. Bender [12,13] has discussed about a special non-hermitian Hamiltonian. The interesting thing about these set of Hamiltonians is that despite being non-hermitian, in a specific condition, they produce real spectra. The underlying condition they need to follow, is to have a $\mathcal{PT}$- symmetry. In 2015, Deffner and Saxena (2015 *Phys. Rev. Lett. 114* 150601) showed that quantum Jarzynski Equality generalizes to $\mathcal{PT}$- symmetric quantum mechanics in a region of unbroken $\mathcal{PT}$- symmetry. Later Zeng and Yong (2017 *Journal of Phys. Commun. 1* 031001) extended this work to Crook's fluctuation theorem. In this paper, we consider a Quantum open system which is described by a $\mathcal{PT}$ – symmetric non-Hermitian Hamiltonian. The correlation between the system and bath is weak so that we can ignore dissipation [19]. We will show that Jarzynski Equality and Crook's fluctuation theorem still hold true.

## 2. Non-Hermitian Hamiltonians with PT symmetry

According to the earlier conviction, in quantum mechanics, a Hamiltonian to give real spectra must follow the condition of Hermiticity. The Hermiticity of a Hamiltonian can be expressed mathematically as

$$H = H^\dagger \qquad (1)$$

Here † signifies matrix transposition followed by complex conjugation. Once we have this relationship, it can be easily shown that the Eigen values of H are real signifying real spectra. Up until the last decade, it was thought that Hermiticity is a necessary condition for a Hamiltonian in quantum mechanics to have real spectra. But recently C. M. Bender [12,13] showed that even if the Hamiltonian is non-Hermitian, it can have real energy spectrum provided H has an unbroken space-time (PT – symmetry) symmetry. In this case (1) can be replaced by the following expression

$$H = H^{PT}$$

Here P stands for the space-reflection operator or parity operator and T stands for time-reversal operator. The conditions followed by P and T operators are as follows [13]

$$P \hat{x} P = -\hat{x} \qquad P \hat{p} P = -\hat{p}$$
$$T \hat{x} T = \hat{x} \qquad T \hat{p} T = -\hat{p}$$

Here $\hat{x}$ and $\hat{p}$ are position and momentum operators respectively. Also
$$T i T = -i$$
Here i is the complex number.

Since we don't have the property of Hermiticity any more, many properties of quantum mechanics need to be modified [21,12,13,18]. The new bra-ket relationship for non-Hermitian quantum mechanics can be expressed as [21,12,13,18]

$|a\rangle \leftrightarrow \langle Ba| = \langle a|B^\dagger = \langle a|B$

Here B is a Hermitian operator such that $B = B^\dagger$

And hence the normalization condition can be expressed as $\langle a|B|a\rangle = 1$. The completeness can be expressed as $\sum_i |a_i\rangle\langle a_i|B = 1$.

The quantum dynamical equation, in the context of non-Hermitian formalism, can be written as [13],

$$i\hbar \frac{\partial |\psi\rangle}{\partial t} = (H(t) + A(t))|\psi\rangle$$

In the above equation, H(t) is the time dependent non-Hermitian Hamiltonian and A(t) is the time dependent gauge field term. The above modified equation preserves the unitarity of the non-Hermitian system.

## 3. Quantum channels

In this section, we will define Quantum Channel. The material of this section is borrowed from [17].

Let's $\mathcal{H}$ be the Hilbert space and $\mathcal{L}(\mathcal{H})$ is the space of linear operators on the Hilbert space $\mathcal{H}$. Now let us consider two Hilbert spaces $\mathcal{H}_A$ and $\mathcal{H}_B$. $\Phi$ is a linear map between $\mathcal{L}(\mathcal{H}_A)$ and $\mathcal{L}(\mathcal{H}_B)$ such that $\Phi: \mathcal{L}(\mathcal{H}_A) \to \mathcal{L}(\mathcal{H}_B)$. For all $X \in \mathcal{L}(\mathcal{H}_A)$,

The linear map can be written as

$$\Phi(X) = \sum_u K_u X K_u^\dagger \qquad (2)$$

Here $K_u$ is the kraus operator. The adjoint of this linear map is written as

$$\Phi^\dagger(Y) = \sum_u K_u^\dagger Y K_u$$

In case of open quantum system, the dynamics is represented by density matrix rather than wave function. The density operator can be written as [20]

$$\rho = \sum_i w_i |i\rangle\langle i|$$

where the system is in state $i$ with probability $w_i$. For a pure state, $\rho$ can be written as [20]

$$\rho = |i_{pure}\rangle\langle i_{pure}|$$

We can use the formalism of linear map mentioned above to establish the relationship between the density matrices in $\mathcal{L}(\mathcal{H}_A)$ and $\mathcal{L}(\mathcal{H}_B)$. If $\rho_A$ is the input then $\Phi(\rho_A)$ is the output. The output density matrix can be written as [17]

$$\rho_B = Tr(\Phi(\rho_A))^{-1} \Phi(\rho_A)$$

from (2), it can easily be shown that

$$\Phi(1_A) = \sum_u K_u K_u^\dagger$$

Or $\quad \Phi(1_A) = 1_B$

In this discussion we are considering a unital quantum channel.

# 4. Jarzynski Equality in non-Hermitian (unbroken $\mathcal{PT}$-symmetric region) quantum open system

For classical system, Jarzynski equality can be written as [1,2]

$$\langle e^{-\beta W} \rangle = e^{-\beta \Delta F}$$

If we consider the correlation between the system and the bath is weak, we can assume that only decoherence is present and dissipation can be neglected [19]. Since there is no dissipation, two-point energy measurement scheme can be considered [9-11]. Let's the initial Eigen state be $\varphi_m(t_0)$ and final $\varphi_n(t_f)$. Now considering only decoherence and neglecting dissipation [19], the work done by the external force in the system can be written as,

$$W = E_n(t_f) - E_m(t_0)$$

Now, applying the conditions of $\mathcal{PT}$-symmetric quantum system,

$$\langle e^{-\beta W} \rangle = \sum_{m,n} \frac{e^{-\beta E_m(t_0)}}{Z(t_0)} \langle \varphi_n(t_f) | B_{t_f} \Phi(\rho_A) | \varphi_n(t_f) \rangle \, e^{-\beta(E_n(t_f) - E_m(t_0))}$$

Here $\langle \varphi_n(t_f) | B_{t_f} \Phi(\rho_A) | \varphi_n(t_f) \rangle$ is the transition probability.

$$\langle e^{-\beta W} \rangle = \sum_{m,n} \frac{e^{-\beta E_m(t_0)}}{Z(t_0)} e^{-\beta(E_n(t_f) - E_m(t_0))} \langle \varphi_n(t_f) | B_{t_f} \Phi(|\varphi_m(t_0)\rangle\langle\varphi_m(t_0)| B_{t_0}) | \varphi_n(t_f) \rangle$$

$$= \frac{1}{Z(t_0)} \sum_n e^{-\beta E_n(t_f)} \langle \varphi_n(t_f) | B_{t_f} \Phi(\sum_m |\varphi_m(t_0)\rangle\langle\varphi_m(t_0)| B_{t_0}) | \varphi_n(t_f) \rangle$$

$$= \frac{1}{Z(t_0)} \sum_n e^{-\beta E_n(t_f)} \langle \varphi_n(t_f) | B_{t_f} \Phi(1) | \varphi_n(t_f) \rangle$$

$$= \frac{1}{Z(t_0)} \sum_n e^{-\beta E_n(t_f)} \langle \varphi_n(t_f) | B_{t_f} | \varphi_n(t_f) \rangle$$

$$= \frac{1}{Z(t_0)} \sum_n e^{-\beta E_n(t_f)}$$

$$= \frac{Z(t_f)}{Z(t_0)}$$

$$= e^{-\beta \Delta F}$$

Since, free energy can be written as $F = -\frac{1}{\beta} \ln Z$ and this is the mathematical representation of

Jarzynski Equality.

# 5. Crook's fluctuation theorem in non-Hermitian (unbroken $\mathcal{PT}$-symmetric region) quantum open system

The probability distribution of work can be written [10,11]

$$P_{t_0 t_f}(W) = \sum_{m,n} \delta(W - (E_n(t_f) - E_m(t_0))\, P\,(|\varphi_m(t_0)\rangle)\mathrm{P}\,(|\varphi_m(t_0)\rangle \to |\varphi_n(t_f)\rangle)$$

Here, $P\,(|\varphi_m(t_0)\rangle)$ denotes the probability of the system found in the Eigen state $|\varphi_m(t_0)\rangle$ and $\mathrm{P}\,(|\varphi_m(t_0)\rangle \to |\varphi_n(t_f)\rangle)$ is the transition probability. $P\,(|\varphi_m(t_0)\rangle)$ can be written as [20]

$$P\,(|\varphi_m(t_0)\rangle) = \frac{e^{-\beta E_m(t_0)}}{Z(t_0)}$$

The transition probability for the Hermitian quantum mechanics can be written as

$$\mathrm{P}\,(|\varphi_m(t_0)\rangle \to |\varphi_n(t_f)\rangle) = \langle \varphi_n(t_f)|\Phi(\rho_A)|\varphi_n(t_f)\rangle$$

$$= \langle \varphi_n(t_f)|\Phi(|\varphi_m(t_0)\rangle\langle\varphi_m(t_0)|)|\varphi_n(t_f)\rangle$$

Considering the first pure energy state of the system to be $|\varphi_m(t_0)\rangle$.

Hence the transition probability for non-Hermitian system in unbroken $\mathcal{PT}$- symmetric region can be written as

$$\mathrm{P}\,(|\varphi_m(t_0)\rangle \to |\varphi_n(t_f)\rangle) = \langle \varphi_n(t_f)\left|B_{t_f}\Phi(|\varphi_m(t_0)\rangle\langle\varphi_m(t_0)|B_{t_0})\right|\varphi_n(t_f)\rangle$$

Next, we will follow the same method adopted by Talkner and Hanggi [10,11]. The Fourier transformation of the work distribution is

$$\tilde{P}_{t_0 t_f}(u) = \int dW\, e^{iuW}\, P_{t_0 t_f}(W)$$

$$= \sum_{m,n} e^{iu(E_n(t_f)-E_m(t_0))}\, e^{-\beta E_m(t_0)}\Big/Z(t_0)$$
$$\langle \varphi_n(t_f)\left|B_{t_f}\Phi(|\varphi_m(t_0)\rangle\langle\varphi_m(t_0)|B_{t_0})\right|\varphi_n(t_f)\rangle \quad (3)$$

The time reversed distribution can be written [10,11], setting $\vartheta = u + i\beta$

$$\tilde{P}_{t_f t_0}(\vartheta) = \int dW\, e^{i\vartheta W}\, P_{t_f t_0}(W)$$

$$= \sum_{m,n} e^{i\vartheta(E_m(t_0)-E_n(t_f))}\, e^{-\beta E_n(t_f)}\Big/Z(t_f)$$

$$\langle\varphi_m(t_0)|B_{t_0}\Phi(|\varphi_n(t_f)\rangle\langle\varphi_n(t_f)|B_{t_f})|\varphi_m(t_0)\rangle$$

(4)

Now, using the property of linear map explained above

$$\langle\varphi_n(t_f)|B_{t_f}\Phi(|\varphi_m(t_0)\rangle\langle\varphi_m(t_0)|B_{t_0})|\varphi_n(t_f)\rangle =$$

$$\langle\varphi_n(t_f)|B_{t_f}\sum_u K_u(|\varphi_m(t_0)\rangle\langle\varphi_m(t_0)|B_{t_0})K_u^\dagger|\varphi_n(t_f)\rangle$$

$$=\sum_u \langle\varphi_n(t_f)|B_{t_f}K_u|\varphi_m(t_0)\rangle\langle\varphi_m(t_0)|B_{t_0}K_u^\dagger|\varphi_n(t_f)\rangle$$

$$=\sum_u \langle\varphi_m(t_0)|B_{t_0}K_u^\dagger|\varphi_n(t_f)\rangle\langle\varphi_n(t_f)|B_{t_f}K_u|\varphi_m(t_0)\rangle$$

$$=\langle\varphi_m(t_0)|B_{t_0}\sum_u K_u^\dagger(|\varphi_n(t_f)\rangle\langle\varphi_n(t_f)|B_{t_f})K_u|\varphi_m(t_0)\rangle$$

$$=\langle\varphi_m(t_0)|B_{t_0}\Phi(|\varphi_n(t_f)\rangle\langle\varphi_n(t_f)|B_{t_f})|\varphi_m(t_0)\rangle \qquad (5)$$

Now from (3)

$$\tilde{P}_{t_0 t_f}(u) = \sum_{m,n} e^{iu(E_n(t_f)-E_m(t_0))} e^{-\beta E_m(t_0)}/Z(t_0)$$
$$\langle\varphi_n(t_f)|B_{t_f}\Phi(|\varphi_m(t_0)\rangle\langle\varphi_m(t_0)|B_{t_0})|\varphi_n(t_f)\rangle$$

$$\Rightarrow Z(t_0)\tilde{P}_{t_0 t_f}(u) = \sum_{m,n} e^{iu(E_n(t_f)-E_m(t_0))}$$
$$e^{-\beta E_m(t_0)}\langle\varphi_n(t_f)|B_{t_f}\Phi(|\varphi_m(t_0)\rangle\langle\varphi_m(t_0)|B_{t_0})|\varphi_n(t_f)\rangle$$

(6)

Again

$$\tilde{P}_{t_f t_0}(\vartheta) = \sum_{m,n} e^{i\vartheta(E_m(t_0)-E_n(t_f))} e^{-\beta E_n(t_f)}/Z(t_f)$$
$$\langle\varphi_m(t_0)|B_{t_0}\Phi(|\varphi_n(t_f)\rangle\langle\varphi_n(t_f)|B_{t_f})|\varphi_m(t_0)\rangle$$

$$\Rightarrow Z(t_f)\tilde{P}_{t_f t_0}(\vartheta) = \sum_{m,n} e^{i\vartheta(E_m(t_0)-E_n(t_f))} e^{-\beta E_n(t_f)}$$
$$\langle\varphi_m(t_0)|B_{t_0}\Phi(|\varphi_n(t_f)\rangle\langle\varphi_n(t_f)|B_{t_f})|\varphi_m(t_0)\rangle$$

$$= \sum_{m,n} e^{i(u+i\beta)(E_m(t_0)-E_n(t_f))} e^{-\beta E_n(t_f)}$$

$$\langle \varphi_m(t_0)|B_{t_0}\Phi(|\varphi_n(t_f)\rangle\langle\varphi_n(t_f)|B_{t_f})|\varphi_m(t_0)\rangle$$

$$= \sum_{m,n} e^{iu(E_n(t_f)-E_m(t_0))} e^{-\beta E_m(t_0)}$$
$$\langle \varphi_m(t_0)|B_{t_0}\Phi(|\varphi_n(t_f)\rangle\langle\varphi_n(t_f)|B_{t_f})|\varphi_m(t_0)\rangle \qquad (7)$$

Now considering the condition (5), from (6) and (7), we can say that

$$Z(t_0)\, \tilde{P}_{t_0 t_f}(u) = Z(t_f)\, \tilde{P}_{t_f t_0}(\vartheta)$$

$$\Rightarrow \frac{\tilde{P}_{t_0 t_f}(u)}{\tilde{P}_{t_f t_0}(\vartheta)} = \frac{Z(t_f)}{Z(t_0)}$$

$$\Rightarrow \frac{\tilde{P}_{t_0 t_f}(u)}{\tilde{P}_{t_f t_0}(u+i\beta)} = \frac{Z(t_f)}{Z(t_0)}$$

After taking the inverse Fourier transformation

$$\frac{P_{t_0 t_f}(W)}{P_{t_f t_0}(-W)} = \frac{Z(t_f)}{Z(t_0)} e^{\beta W} = e^{\beta(W-\Delta F)}$$

And this is nothing but the mathematical representation of Crook's fluctuation theorem.

The above derivation is valid when we have unbroken PT- symmetry. In case of broken PT-symmetry the dynamics is no longer unitary [14] and it can be concluded that the crook's theorem and Jarzynski Equality are not valid for the Quantum open system with broken PT-symmetry.

## 6. Conclusion

In the above discussion, it has been shown that both Jarzynski Equality and Crook's Fluctuation Theorem valid in non-Hermitian open quantum system with unbroken PT-symmetry. In this work, the coupling between the system and the Bath is considered to be weak so that we can assume dissipation to be negligible compared to decoherence. A future goal would be to examine the validity of these theorems when the coupling is strong and both dissipation and decoherence are significant.